\documentclass[fleqn,12pt,twoside]{article}
\usepackage{espcrc1}

\usepackage{graphicx}
\usepackage[figuresright]{rotating}

\newcommand{\AmS}{{\protect\the\textfont2
  A\kern-.1667em\lower.5ex\hbox{M}\kern-.125emS}}

\title{The Influence of Reaction Rates on the Final p-Abundances}

\author{W. Rapp\address[MCSD]{University of Notre Dame,\\ 
                 Department of Physics \&
                 Joint Institute of Nuclear Astrophysics,\\
                 225 Nieuwland Science Hall, Notre Dame, 
                 IN 46556, USA}\thanks{WR \& MW are supported 
                 through the 
                 Joint Institute of Nuclear Astrophyisics by NSF-PFC 
                 grant PHY02-16783.},
        M. Wiescher\addressmark[MCSD],
        H. Schatz\address{Department of Physics and Astronomy, National
                Superconducting Cyclotron Laboratory \& Joint Institute
                of Nuclear Astrophysics, Michigan State University,\\  
                East Lansing, MI 48824, USA}\thanks{HS is an 
                Alfred P. Sloan Fellow and is supported by NSF 
                grants PHY 02-16783 (JINA) and PHY01-10253 (NSCL).},
        and
        F. K\"appeler\address {Forschungszentrum Karlsruhe,\\ 
                Institut f\"ur Kernphysik,\\  
                76021 Karlsruhe, Germany}}
       
\begin{document}

\maketitle

\begin{abstract} 
The astrophysical p-process is responsible for the origin of the
proton rich nuclei,which are heavier than iron. A huge network
involving thousands of reaction rates is necessary to calculate
the final p-abundances. But not all rates included in the network
have a strong influence on the p-nuclei abundances.\\ 
The p-process was investigated using a full nuclear reaction network
for a type II supernovae explosion when the shock front passes
through the O/Ne layer. Calculations were done with a multi-layer
model adopting the seed of a pre-explosion evolution of a 25 mass
star. In extensive simulations we investigated the impact of
single reaction rates on the final p-abundances.
The results are important for the strategy of future experiments
in this field.
\end{abstract}

\section{INTRODUCTION}
Almost 100$\%$ of the heavy nuclei are produced by neutron capture processes, because 
of the high Coulomb barrier reduces the impact of charge particale reactions.
In nature, 32 nuclei can be found on the left side of the stable valley from $^{74}$Se 
to $^{196}$Hg, 
which are shielded against neutron capture. For the synthesis of these nuclei the p-process 
has to be estabished. 
The most favored scenario for the p-process are type II supernova (SN) 
explosions. In these environments high temperatures 1.7$<$T$_{9}$$<$3.3 
can be reached for a 
short time when the shock front passes though the O/Ne layers \cite{RAH95}. 
A large reaction network involving thousands of reaction rates is necessary to describe the 
synthesis process for the p-nuclei. In most of the cases the astrophysical rates were calculated 
by means of the Hauser Feshbach model. In experimental efforts only a few of the required 
astrophysical rates have been investigated \cite{ArG03}. 
Recent experiments \cite{SFK98,RKK03} have shown that the present statistical models 
have overestimated the values for $\alpha$-induced astrophysical rates. The aim of this work is 
to look for rates (A$>$57) that have a significant influence of 
the p-abundences for further experimental efforts.

\section{SIMULATIONS}
For the following investigations a nuclear reaction network used for x-ray bursts at 
Michigan State University was extended. The network now contains about 1800 nuclei covering 
all elements from hydrogen to bismuth. 
Simulated was the abundance evolution during a SN explosion (10$^{51}$erg) in 14 different 
O/Ne-layers for a 25 solar mass star during the first second of the explosion.
The temperature and density profiles 
were available as function of time in tabulated form \cite{YoH02}. The seed abundances were taken 
from a pre-SN model \cite{RAH95}. In the calculations the n-, p-, $\alpha$- and their 
inverse rates 
from the NON-SMOKER code \cite{RaT02} were used (Z$>$8) for all other reactions the data 
\cite{Sch02} were included. 
The reaction rates were changed in the frame of a parameter study.\\
{\bf Collective change of rates:} All n-induced and their inverse rates (A$>$57) were 
multiplied by a factor of three and divided, respectively. In each case the p-abundances 
were calculated 
and the overproduction factors were compared to the result based on the presently accepted HF-rates. The p- and $\alpha$-induced rates were treated in the same way.\\
{\bf Change of p-nuclei rates:} All n-induced and their
inverse reactions including p-nuclei were changed going to the 
neutron rich side and to the neutron deficient side of the stable valley, 
respectively. 
The same procedure was done for all p-induced and their
inverse rates, going to the proton rich and proton deficient side respectively 
and for $\alpha$-rates as well. 
The simulations were performed analog to the collective change of rates. \\
{\bf Branchings:} In the integrated reaction flux all branchings were detected (A$>$57) when 
the second strongest flux was greater than 20$\%$ of the main flux. The branching ratios 
were changed and the influence on the p-abundances was determined. 

\section{RESULTS (Selected)}
\subsection{Collective change of rates}
Figure 1 is showing the ratio of the p-abundances calculated with changed rates
divided by the abundances simulated with the unchanged rates.
A collective change of n-induced reactions has an influence on the whole mass range, espacially
in the lower A$<$80 and upper A$>$150 mass ranges.
The influence of the p-induced reactions is only at lower mass range A$>$110.
In contrast to the p-induced reactions, $\alpha$-induced reactions showed a big influence on 
the p-abundances at high mass A$>$140, with a exception the $^{92}$Mo($\alpha$,$\gamma$)-reaction.

\subsection{Change of p-nuclei rates:} 
For n-induced reactions the change in individual rates were much less 
sensitive compared to the collective change of the rates. 
The change of (n,$\gamma$)-rates on the p-nuclei showed almost no influence 
while the change of the ($\gamma$,n)-reactions showed a sensitivity. 
The ($\gamma$,p)-rates determine significantly the p-abundances 
of $^{74}$Se, $^{78}$Kr, $^{84}$Sr, $^{92}$Mo, and $^{96}$Ru (A$<$100). Proton capture rates 
have almost no influence on the p-abundances.
The comparision of the abundances calculated with the collective and the individually change 
of p-induced rates   
suggest that for proton involved reactions the abundances of $^{78}$Kr, $^{92}$Mo, and $^{96}$Ru are 
only sensitive to the photodisintegration of these nuclei.

\subsection{Branchings:} 
More than 200 branches were detected in the integrated reaction flux. 
Most of the branchings showed no influence on the p-abundances. But for A$>$140 a set of 
($\gamma$,$\alpha$)- and ($\gamma$,n)-branches were sensitive for p-abundances. 
These branching rations were detected in a small temperature window
 2.4$<$T$_{9}$$<$2.8. 
In this temperature window, the reaction flow is so dynamic 
that a change of these branchings showed an influence on several 
p-nulei. The changes in the p-abundances were in most of the cases less than 30$\%$ in the 
particular layer investigated, when increasing the second strongest branching reaction by a 
factor of 3. 
But the result of using a wrong potential to calculate the astrophysical reaction rates 
can be seen in Fig. 1. 
Further experiments should measure $\alpha$-induced reactions in the mass 
range A$>$150, because there they have a significant influence on the p-abundances 
and there are no experimental data at all.

\section{ACKNOWLADGEMENT}
We could like to thank T. Yoshida form the Astronomical Data Analysis Center, 
National Astronomical Observatory in Osawa, Mitaka, Tokyo, (Japan) 
for providing us with the temperature and density profiles. Thanks to M. Rayet from
University of Brussels (Netherlands) as well, for sending us the s-process seed and
we are much obliged to F.-K. Thielemann from the Deptartment of Physics and Astronomy, 
University of Basel (Switzerland) for his pioneering work on the 
reaction network. This project is supported through the Joint Institute of 
Nuclear Astrophysics by NSF-PFC grant PHY02-16783.

\newpage
\begin{figure}[tp]
\begin{center}  
\includegraphics*[width=9cm]{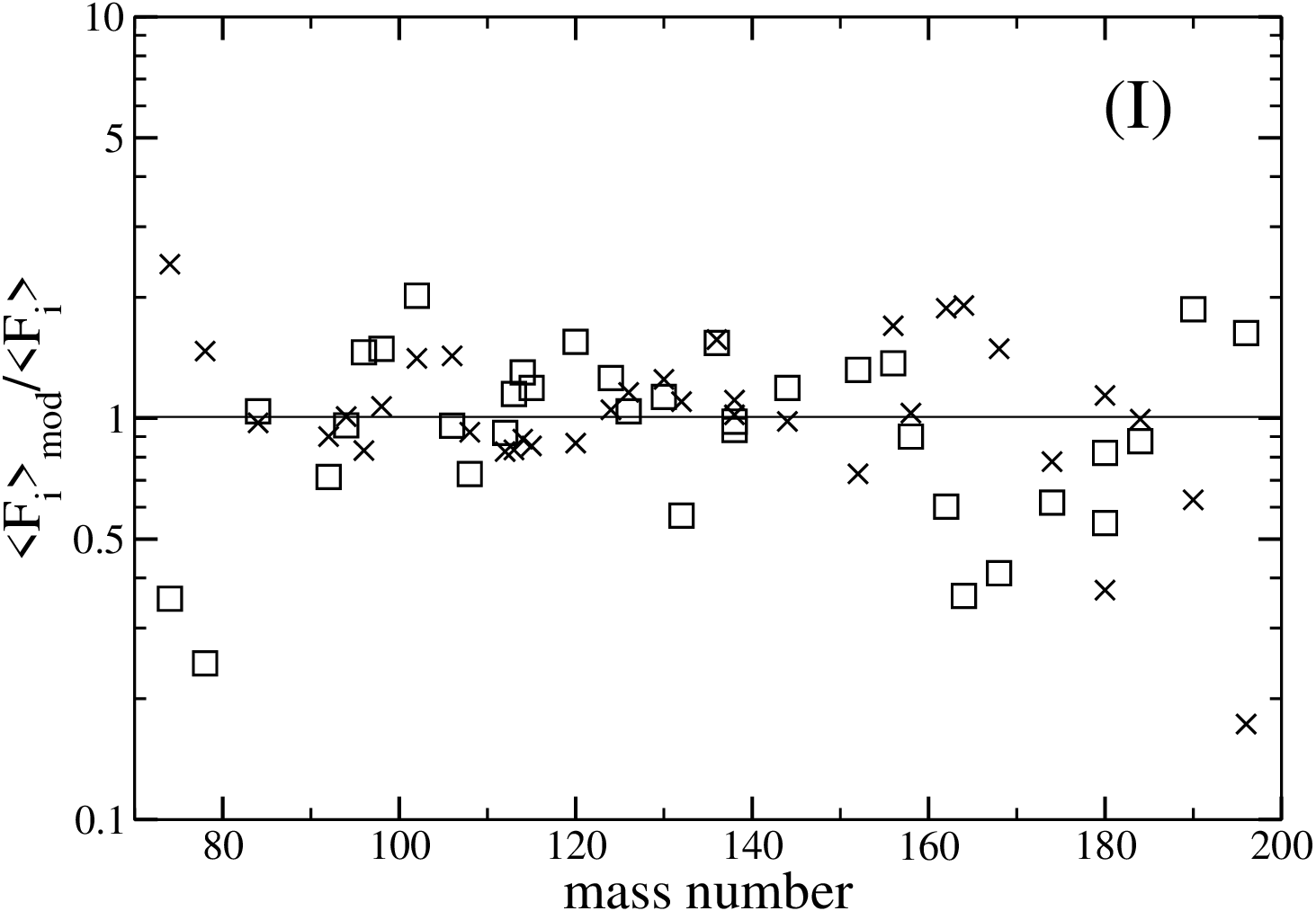}
\end{center} 
\begin{center}  
\includegraphics*[width=9cm]{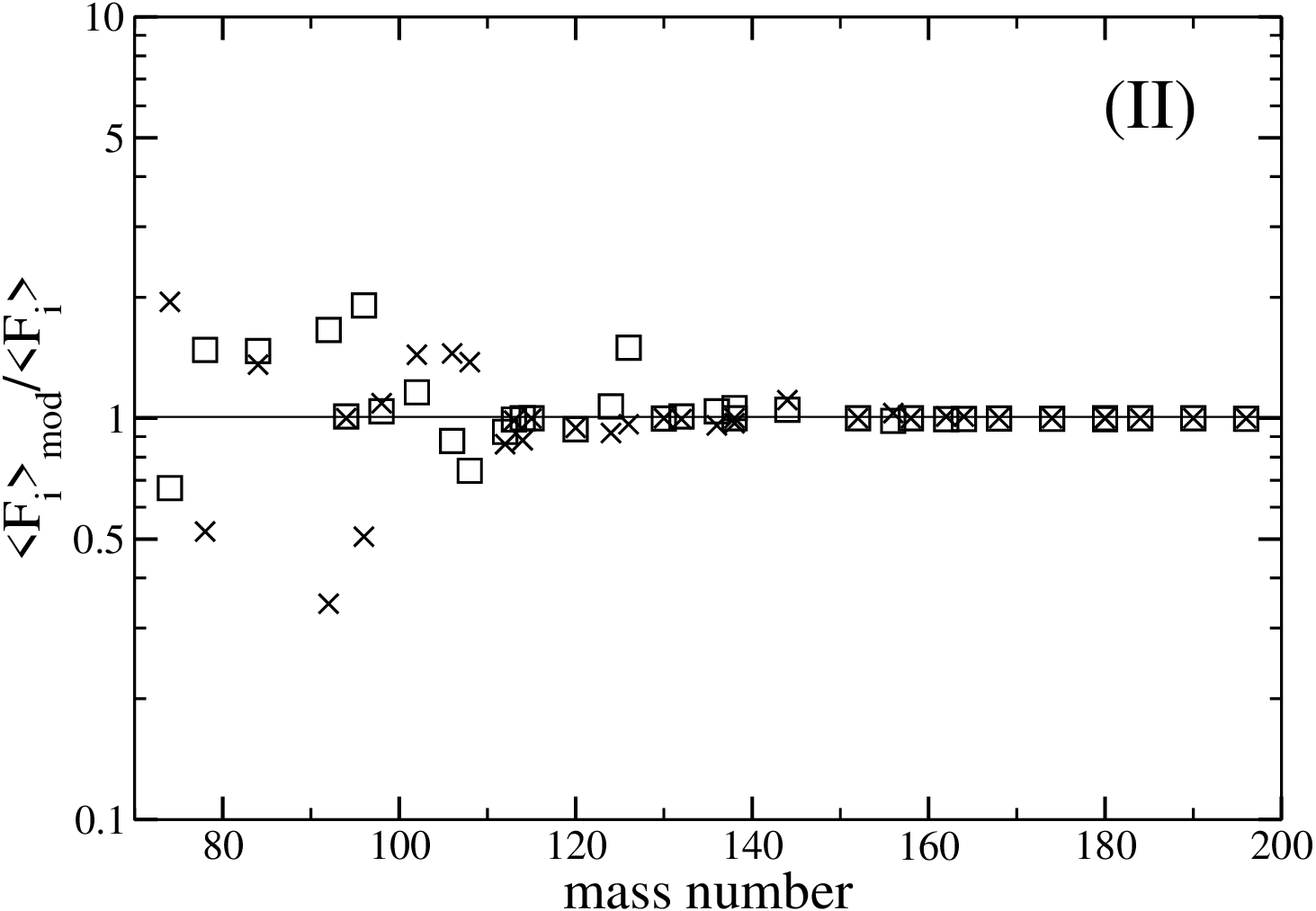}
\end{center}
\begin{center}  
\includegraphics*[width=9cm]{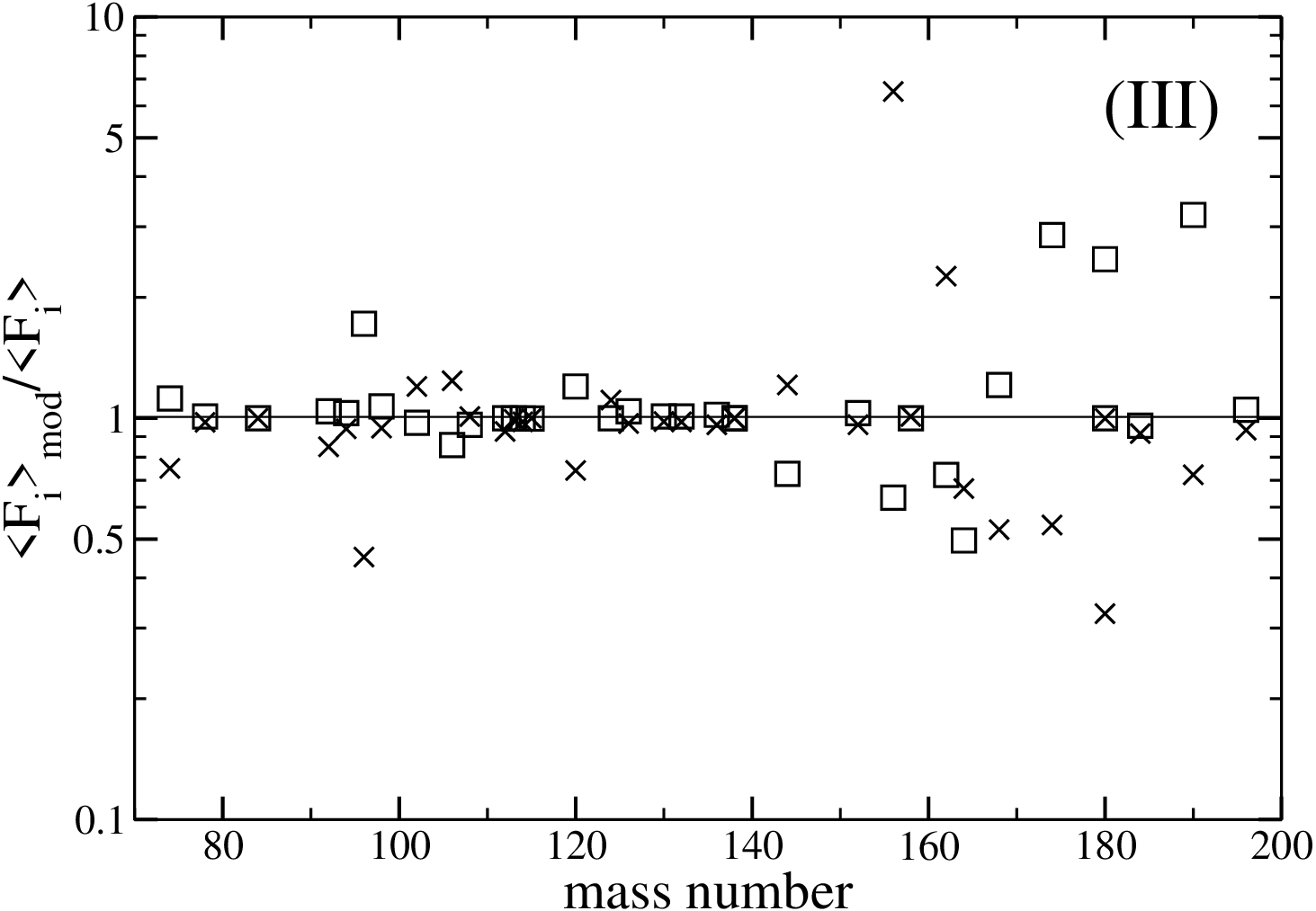}
\caption{Shown are the ratios of the p-abundances calculated with changed rates
divided by the abundances based on the presently accepted HF-rates 
(squares factor 1/3; cruxes factor 3) 
for all n-induced (I), p-induced (II), and $\alpha$-induced reactions (III).}
\end{center}  
\end{figure}

\end{document}